\documentstyle[graphicx]{aipproc}

\title{Wavelet Analysis and Lognormal Distributions in GRBs}
\author{Kevin J. Hurley, Brian McBreen, Fergus Quilligan, Matt Delaney and 
Lorraine Hanlon}
\address{Physics Dept., University College, Dublin 4, Ireland.}

\begin{document}

\maketitle

\begin{abstract}
  A wavelet analysis has been performed on 80 intense gamma-ray bursts
  (GRBs) from the BATSE 3B catalog with durations longer than 2
  seconds.  The wavelet analysis applied novel features developed for
  edge detection in image processing and this filtering process was
  used to extract a fit to the irregular GRB profile from the
  background.  A straightforward algorithm was subsequently used to
  identify statistically significant peaks in this profile. The areas
  and FWHM of 270 peaks that were characterised as isolated were found
  to be consistent with lognormal distributions. The distribution of
  time intervals between peak maxima for all 963 identified peaks in
  the GRBs is also presented.
\end{abstract}

\section*{Introduction}

The recent significant discoveries of fading x--ray emission
\cite{t-p01:costa.e:97-nat-387-783} have resulted in the first
identification of radio and optical counterparts to GRBs
\cite{t-p01:frail.d:97-iauc-6662,t-p01:vanparadijs.j:97-nat-386-686}. The optical
spectrum of one counterpart (GRB\,970508) has yielded a value of
z=0.835 \cite{t-p01:metzger.mr:97-nat-387-878}.

GRBs produce an immense amount of energy ($\sim 10^{51}$\,ergs) which
must be emitted by a medium with highly relativistic velocities
($\gamma > 100$). Most GRBs are highly variable with a variability
scale much smaller than their overall durations and there is
considerable debate as to the cause of the variability, which may be
due to internal or external shocks \cite{t-p01:chiang.j:97-astroph-9708035,%
t-p01:panaitescu.a:97-astroph-9703187,t-p01:sari.r:97-astroph-9701002}.

GRBs generally have complex time profiles and previous work
\cite{t-p01:mcbreen.b:94-mn-271-662} has indicated that some parameters are
consistent with lognormal distributions e.g. durations of GRBs and
time intervals between peaks. BATSE bursts with multiple peaks were
also analysed and the applicability of lognormal distributions in GRBs
was confirmed\cite{t-p01:li.h:96-apj-469-l115}. Other techniques have also
been used to profile the time structure of GRBs
\cite{t-p01:mitrofanov.i:95-ass-231-103,t-p01:norris.jp:96-apj-459-393,t-p01:stern.be:96-apj-469-l109}.  The
results presented here on a large sample of BATSE GRBs were obtained
using wavelets.

\section*{Data Preparation}

The data was obtained from the BATSE 3B catalogue. A
subset of this catalogue was selected based on criteria used by Norris
and co-workers \cite{t-p01:norris.jp:96-apj-459-393}, which gave a sample of
87 long bright bursts ($T_{90}>$2 s and $P_{256{\rm ms}} > 4.6$ ph
cm$^{-2}$ s$^{-1}$). The start and end times for each burst were
identified by eye and a margin of 10 seconds added. Two sections of
duration 30 s, one at -50 s from the start and the other at +20 s from
the end were used to provide a linear interpolation for background
subtraction.

From the original sample of 87 bursts, a smaller sample of 80
background subtracted GRBs suitable for denoising (those with no
data-gaps) were processed with a multiscale edge detection wavelet
denoising routine \cite{t-p01:mallat.sg:92-i3etpami-7-710,t-p01:mallat.sg:92-i3eit-38-617,t-p01:young.ca:95-ass-231-119,t-p01:daubechies.i:92-b-tlow}.

\begin{figure}[tbp]
  \begin{center}
    \leavevmode
    \includegraphics[angle=270, width=\textwidth]{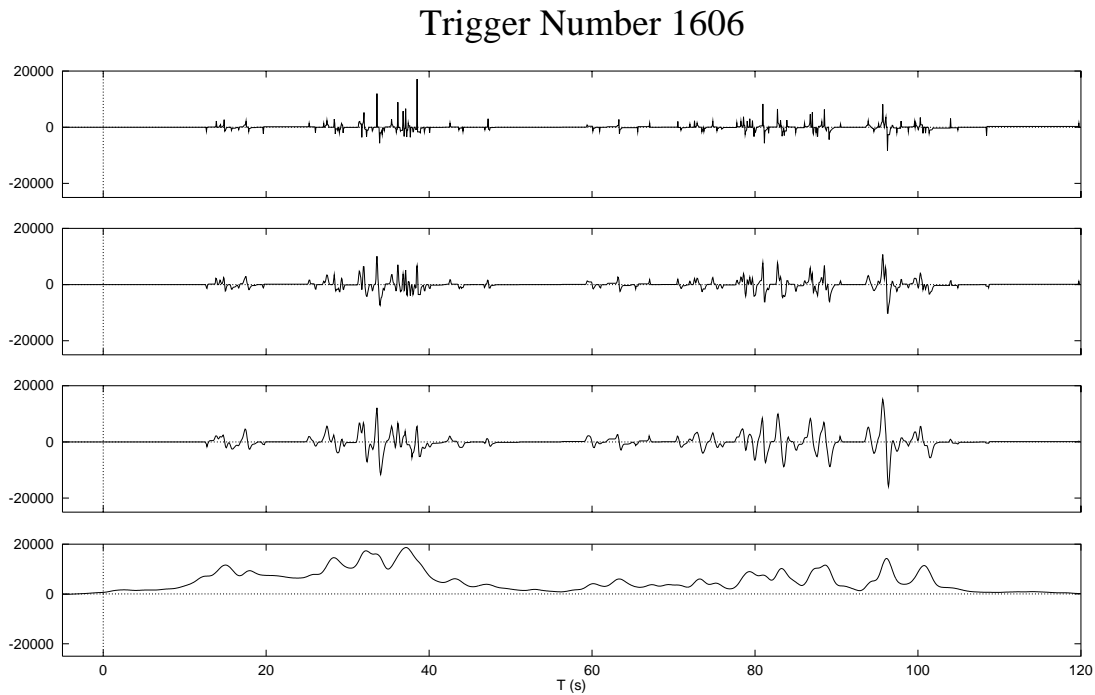}
  \end{center}
\caption[]{\label{t-p01:fig:scales} 
  A wavelet transformation representation of GRB 920513 (BATSE trigger
  1606) on three scales and the remaining low frequency signal.  Note
  that the extreme values in the transform correspond to the edges in
  the signal and the signal can be reconstructed from just these
  extreme values\protect{\cite{t-p01:mallat.sg:92-i3etpami-7-710}}. This fact
  allows the removal of noise structure by removing the maxima and
  minima corresponding to the noise edges before the signal
  reconstruction\cite{t-p01:mallat.sg:92-i3eit-38-617}.}
\end{figure}

%Mallat and Zhong \cite{t-p01:mallat.sg:92-i3etpami-7-710} developed an
%algorithm for allowing the reconstruction of a signal given just the
%maxima and minima of the wavelet transforms across a set of scales
%(along with the low frequency signal remaining), as in
%Figure~\ref{t-p01:fig:scales}. In a companion paper Mallat and Hwang
%\cite{t-p01:mallat.sg:92-i3eit-38-617} introduce a technique for eliminating
%noise from this representation, hence allowing reconstruction of a
%relatively noise-free version of the original signal. It is this
%technique, the multiscale edge-detection algorithm, which forms the
%cornerstone for the denoising method used here.

Following on the wavelet denoising of the original GRB, a method was
devised to identify the peak structures within the burst.  Starting at
the beginning of the burst, each peak was examined for minima on both
sides which were separated from the maximum by more than a chosen
significance level. If the search for minima failed on this peak it
was rejected and the search continued. The algorithm was designed so
that the lowest minima were selected for each maxima. The analysis was
extended to allow the identification of peaks which are well separated
or isolated. The fraction of the total height which was above the
higher minimum for every peak was calculated and then a threshold was
applied on this fraction above which peaks were considered isolated.

\section*{Results}

\begin{figure}[tb]
  \begin{center}
    \leavevmode \includegraphics[angle=270,
    width=0.75\textwidth]{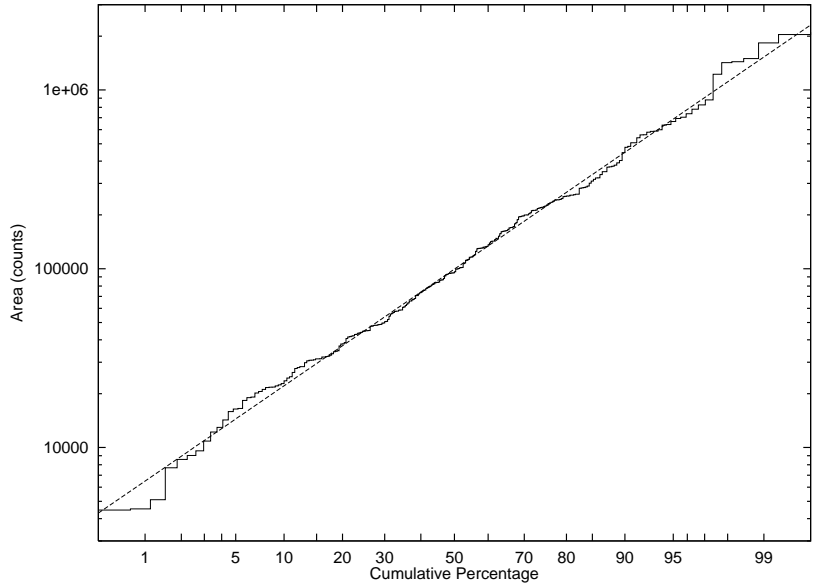}
  \end{center}
    \caption{%
      The distribution of the areas of 270 peaks that were isolated at
      the 66\% level (see text). The dashed line is a very acceptable
      lognormal fit to the data (see
      Table~\ref{t-p01:tab:summary}).}
    \label{t-p01:fig:area}
\end{figure}

\begin{figure}[tb]
  \begin{center}
    \leavevmode
    \includegraphics[angle=270, width=0.75\textwidth]{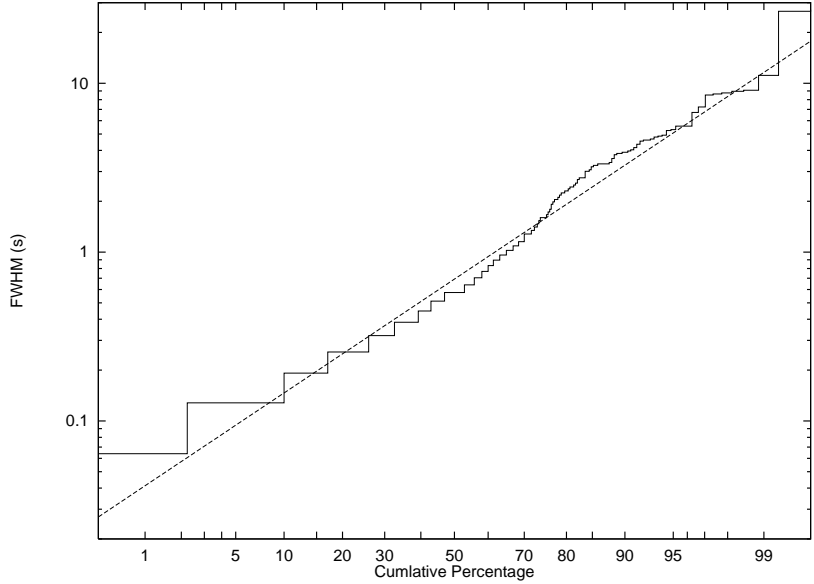}
  \end{center}
    \caption{%
      The FWHM distribution for the same peaks as in
      Figure~\ref{t-p01:fig:area}. Again the dashed line is a lognormal fit
      to the data.}
    \label{t-p01:fig:fwhm}
\end{figure}

\begin{figure}[tb]
  \begin{center}
    \leavevmode
    \includegraphics[scale=.37,angle=180]{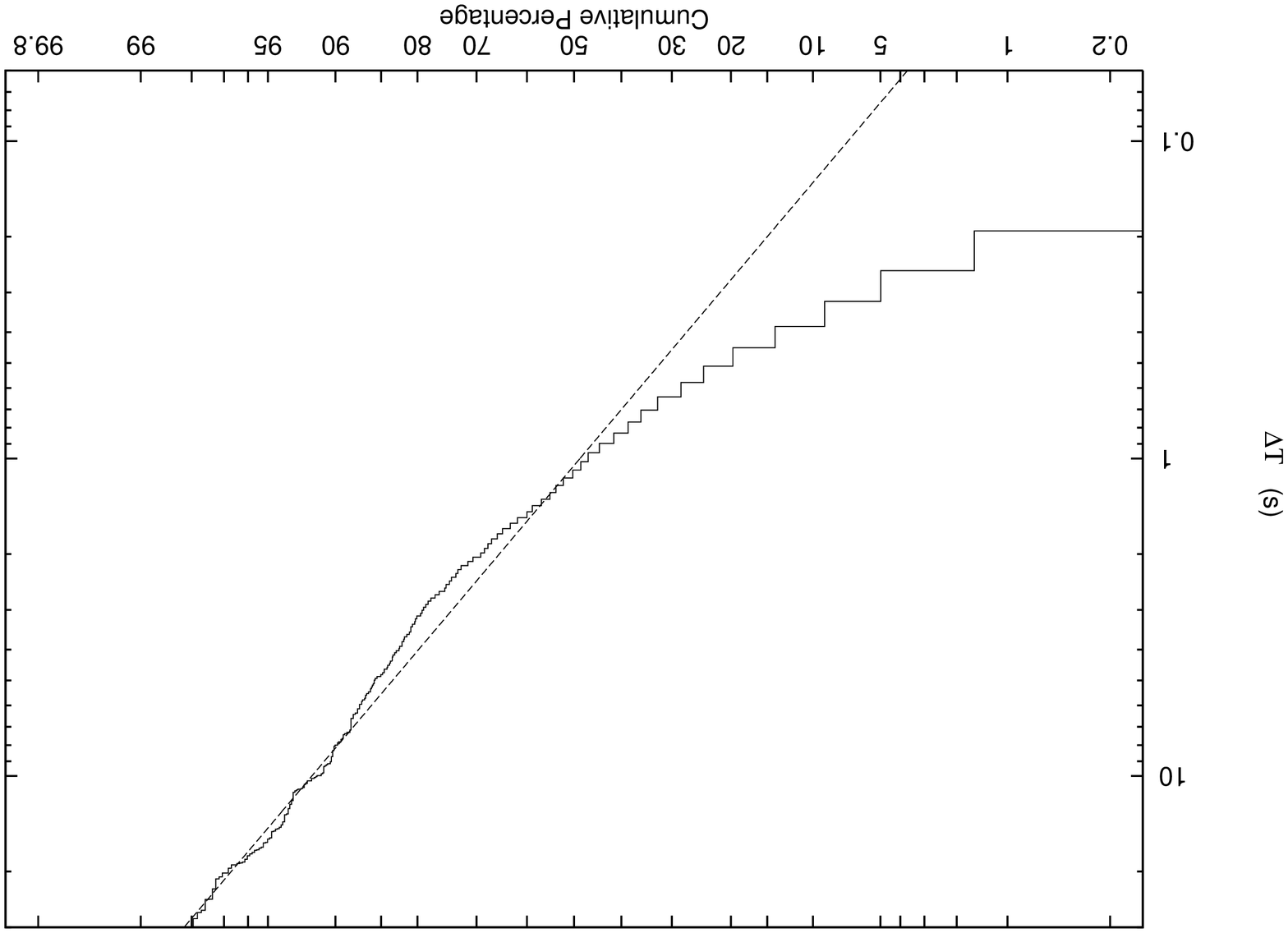}
  \end{center}
  
  \caption[]{%
    The distribution of time intervals between the 963 peaks
    identified in 80 GRBs. The dashed line indicates a candidate
    lognormal fit. The departure of the fit at short
    time intervals is strongly influenced by the limited resolution of
    the data\cite{t-p01:hurley.kj:97-inprep}.}
  \label{t-p01:fig:dt}
\end{figure}

Having identified peaks in the GRB time profiles, certain key
characteristics of the data were examined. In particular the time
interval between peaks ($\Delta {\rm T}$), the area under peaks
(measured as total counts above background while the signal is above
the height threshold for isolation) and the full width at half maximum
(FWHM) were investigated. The distributions of each of these
properties (Figures 2 and 3) was tested for compatibility with the lognormal
distribution. The results of the statistical tests are presented in
Table~\ref{t-p01:tab:summary}. It should be noted that the peak area and FWHM are quite compatible with the lognormal distribution.

\section*{Conclusions}

\begin{table}[b]
  \caption{% 
Summary of goodness of fit tests for lognormal
distributions. $\chi^2_{2\sigma}$ is the $2\sigma$ confidence limit of
the reduced $\chi^2$ value ($\chi^2/\nu$) with $\nu$ degrees of
freedom.}  

\label{t-p01:tab:summary}
      \begin{tabular}{lddddd}
        \multicolumn{1}{c}{Dist.} &\multicolumn{1}{c}{Figure} &\multicolumn{1}{c}{$\mu$} & \multicolumn{1}{c}{$\sigma$} & 
        \multicolumn{1}{c}{$\chi^2_\nu/\nu$} & \multicolumn{1}{c}{$\chi^2_{2\sigma}$} \\
        \tableline
        Area & 2 & 11.51 & 1.17 & 1.033 & 1.326\\
        FWHM & 3 & -0.368 & 1.212 & 1.276 & 1.306\\
        $\Delta{\rm T}$ & 4 & 0.05 & 1.0 & 4.119 & 1.326\\
      \end{tabular}

\end{table}

A comprehensive profile of the temporal properties of intense GRBs has
now been provided by the wavelet method. Many properties of GRBs are
consistent with lognormal distributions:
\begin{itemize}
\item The bimodal duration distribution of GRBs is consistent with two
  lognormal distributions\cite{t-p01:mcbreen.b:94-mn-271-662}.
\item The area of the peaks is consistent with a lognormal distribution.
\item The FWHM of the peaks is consistent with a lognormal distribution.
\item The time intervals between peaks in GRBs as measured with this 
technique does not seem to be consistent with a lognormal distribution 
but important selection effects have not been considered.
\end{itemize}

These results are in good agreement with previous work
\cite{t-p01:mcbreen.b:94-mn-271-662,t-p01:li.h:96-apj-469-l115}.  These results provide an alternative 
description of GRB profiles to that of the peak-aligned stretched 
exponentials \cite{t-p01:mitrofanov.i:95-ass-231-103,t-p01:norris.jp:96-apj-459-393,t-p01:stern.be:96-apj-469-l109}. The data presented here also
provide considerable constraints on models that attribute the
variability to interactions with an external medium or to an internal
origin that reflects the activity of the central engine.

\end{document}